\documentstyle[titlepage,12pt]{article}
\def\half{\frac{1}{2}}

\def\be{\begin{equation}}

\def\ee{\end{equation}}
\def\rq#1{(\ref{eq#1})}

\def\to{\rightarrow}
\def\gt{\tilde{g}}
\def\Rt{\tilde{R}}
\newcount\sectionnumber
\def\sect
{\global\equationnumber=0
\global\advance\sectionnumber by 1
\the\sectionnumber . }
\newcount\equationnumber
\def   \num
{\eqno{\global\advance\equationnumber by 1
\left(\the\sectionnumber .\the\equationnumber \right)}}
\begin{document}
\title{The $D\to 2$ Limit of General Relativity}

\author{R.B. Mann and S.F. Ross\\
        Department of Physics\\
        University of Waterloo\\
        Waterloo, Ontario\\
        N2L 3G1}

\date{August 11, 1992\\
WATPHYS TH 92/06}

\maketitle

\begin{abstract}
A method for taking the $D\to 2$ limit of $D$-dimensional general relativity
is constructed, yielding a two-dimensional theory
which couples gravitation to conserved stress-energy.
We show how this theory is related to those obtained via an alternative
dimensional reduction approach.
\end{abstract}

The study of gravity in two spacetime dimensions has been of considerable
interest for several years
\cite{Jackiw,Arnold,MST,semi,SharTomRobb,MSW,EDW,symbh}. Such theories can
have quite a rich and interesting structure which reproduces qualitatively
much that is found in general relativity, ({\it e.g.} black holes, FRW-type
cosmologies, gravitational collapse) even though they are mathematically
much simpler. This simplicity makes them both a useful pedagogical tool and
an interesting arena for the study of quantum gravitational effects.

The topological character of the Einstein-Hilbert action in $D=2$
dimensions has led theorists to construct such gravitational theories in a
manner which circumvents the Einstein equations \cite{Jackiw,EDW,Rsimp}.
Indeed, the triviality of Einstein's equations in two spacetime
dimensions seemingly indicates that $D=2$ general relativity does not make
sense. We wish to point out here that this is not the case.  Specifically,
we show that (at least formally), one can take the $D\to 2$ limit of
Einstein's equations. This yields a 2D theory of gravity previously
considered in the literature \cite{Arnold,MST,semi,SharTomRobb} in which
gravitation is generated by stress-energy and stress-energy is in turn
acted upon by gravitation, just as in $D>2$ general relativity. We then
elucidate via a dimensional-reduction argument the connections between this
approach and other  2D theories \cite{Jackiw,EDW}.

We begin with the gravitational action in $D$ dimensions
\be
S = \frac{1}{\kappa_D} \int d^Dx \sqrt{-g}R + {\cal L}^{(D)}_M,
\label{eq10}
\ee
where ${\cal L}^{(D)}_M$ is the $D$-dimensional matter Lagrangian. Since the
Einstein tensor goes to zero like $(1-\frac{D}{2})$  when $D \to 2$, we can
extract non-trivial dynamics in this limit by assuming that the
gravitational coupling constant $\kappa_D$ also vanishes like $(1-
\frac{D}{2})$ in this limit \cite{MannGRG}.

Also, we may subtract
a term $\int d^Dx \sqrt{-\gt}\Rt/\kappa_D$ without changing the result in
two dimensions if it becomes a total derivative as $D \to 2$, that is, if
\be
\lim_{D \to 2}\frac{1}{\kappa_D}\tilde{G}_{\mu\nu} = 0. \label{eq11}
\ee
We assume this
is the case for some metric $\gt_{\mu\nu}$ conformal to $g_{\mu\nu}$ (i.e.,
$\gt_{\mu\nu} = e^{\Psi} g_{\mu\nu}$). Then
\cite{Wald}
\be
\Rt = e^{-\Psi} ( R - (D-1)g^{\mu\nu}\Psi_{;\mu\nu} -
\frac{1}{4}(D-2)(D-1) g^{\mu\nu}\Psi_{;\mu}\Psi_{;\nu}).    \label{eq13}
\ee
Since $\sqrt{-\gt} = e^{D\Psi /2}\sqrt{-g}$, the action
\be
S = \frac{1}{\kappa_D}\int d^Dx (\sqrt{-g}R-\sqrt{-\gt}\Rt)
+ {\cal L}^{(D)}_M \label{eq12}
\ee
becomes
\begin{eqnarray}
S &=& \frac{1}{\kappa_D}\int d^D x \sqrt{-g}\left( R -
e^{(\frac{D}{2}-1)\Psi} [R - (D-1) g^{\mu\nu}\Psi_{;\mu\nu} \right.
\nonumber \\ &-& \left. \frac{1}{4}(D-2)(D-
1)g^{\mu\nu}\Psi_{;\mu}\Psi_{;\nu}]\right) + {\cal L}^{(D)}_M. \label{eq14}
\end{eqnarray}
The quantity of interest is the limit of this action as $D \to 2$, so we
may expand the exponential and discard terms of order $(\frac{D}{2}-1)^2$. The
action in this limit is then
\begin{eqnarray}
S& = & \frac{1}{\kappa_D} \int d^Dx\sqrt{-g}
\left((D-1)g^{\mu\nu}\Psi_{;\mu\nu}
+\frac{1}{4}(D-2)(D-1)g^{\mu\nu}\Psi_{;\mu}\Psi_{;\nu} \right. \nonumber \\
& - &\left. (\frac{D}{2}-1)\Psi R+(\frac{D}{2}-1)(D-1)\Psi
g^{\mu\nu}\Psi_{;\mu\nu} \right) + {\cal L}^{(2)}_M + O(\frac{D}{2}-1)^2.
\label{eq15}
\end{eqnarray}
We discard the leading term, as it is a total derivative, and integrate the
final term by parts to obtain
\be
S = \frac{1-\frac{D}{2}}{\kappa_D} \int d^Dx\sqrt{-g} (\Psi R + \half(D-1)
g^{\mu\nu}\Psi_{;\mu}\Psi_{;\nu})+{\cal L}^{(2)}_M + O(\frac{D}{2}-1)^2.
\label{eq16}
\ee

We now wish to choose $\kappa_D$ so as to obtain agreement with the
Newtonian theory in two dimensions. This may be done by taking $\lim_{D \to
2} \kappa_D = (1-\frac{D}{2})8\pi G$ \cite{MannGRG}, and we thus obtain the
limit of the action \rq{12} as $D \to 2$:
\be
S = \frac{1}{8 \pi G} \int d^Dx\sqrt{-g} (\Psi R + \half g^{ab}
\Psi_{;a}\Psi_{;b}) + {\cal L}^{(2)}_M. \label{eq17}
\ee
This action yields the field equation and conservation law
\be
R = 8\pi GT,  \qquad  T^{ab}_{;\ b}=0   \label{eq2}
\ee
along with an auxiliary equation for $\Psi$
\be
\half(\Psi_{;a}\Psi_{;b}-\half g_{ab}\Psi_{;c}\Psi^{;c})+
g_{ab}g^{cd}\Psi_{;cd}-\Psi_{;ab} = 8\pi GT_{ab}.
\label{eq3}
\ee
Note that the classical evolution \rq2 of the gravity/matter system is
independent of the evolution of $\Psi$, although the converse is not
true \cite{MannGRG}.  This theory has already been studied in some detail
\cite{Arnold,MST,semi,SharTomRobb,match}, and its classical and
semi-classical properties exhibit remarkably strong similarities to general
relativity.  It reduces to the lineal theory
of gravity studied by Jackiw \cite{Jackiw} when the stress energy tensor is
taken to be a constant ($T_{ab}\sim\Lambda g_{ab}$).

The condition \rq{11} on the metric $\gt_{\mu\nu}$ may be translated into a
condition on $\Psi$ by using the conformal relation
$\gt_{\mu\nu} = e^{\Psi} g_{\mu\nu}$, which implies
\begin{eqnarray}
\tilde{G}_{\mu\nu}& = &G_{\mu\nu}+(\frac{D}{2}-1)\left(\half(\Psi_{;\mu}
\Psi_{;\nu}-\half g_{\mu\nu}g^{\rho\sigma}\Psi_{;\rho}\Psi_{;\sigma}) \right.
\nonumber \\
& + & \left. g_{\mu\nu}g^{\rho\sigma}\Psi_{;\rho\sigma}-\Psi_{;\mu\nu}\right)+
\half(\frac{D}{2}-1)^2g_{\mu\nu}g^{\rho\sigma}\Psi_{;\rho}\Psi_{;\sigma},
\label{eq18}
\end{eqnarray}
and $G_{\mu\nu} = \kappa_D T_{\mu\nu}$, which follows from the action \rq{10}.
Thus, \rq{11} is equivalent to
\be
\lim_{D \to 2} T_{\mu\nu}-\frac{1}{8\pi G}\left(\half(\Psi_{;\mu}\Psi_{;\nu}-
\half g_{\mu\nu}g^{\rho\sigma}\Psi_{;\rho}\Psi_{;\sigma})+g_{\mu\nu}
g^{\rho\sigma}\Psi_{;\rho\sigma}-\Psi_{;\mu\nu}\right)=0.
\label{eq19}
\ee
The $\Psi$ field equation \rq3 guarantees that the quantity appearing in
the limit \rq{19} vanishes for $D=2$, so we need only require that this
quantity ``vary'' smoothly for $D \to 2$. That is, we require that $\Psi$
{\em as a function of $D$} must be $C^2$ in some neighborhood of $D=2$.

An alternate approach for obtaining 2D actions from 4D general relativity
is via dimensional reduction \cite{FroStrom}.
Consider a spherically symmetric metric in four dimensions
\be
ds^2 = e^{\sigma(r)} g_{ab}dx^a dx^b+ {\cal D}(r)(d\theta^2+\sin^2\theta
d\phi^2),
\label{eq6}
\ee
where latin indices run from $0$ to $1$, $g_{ab}(x^a)$ is the metric on the
two-dimensional submanifold, and $r=r(x^a)$ is a scalar field on this
submanifold. The Ricci scalar for this metric may be written as
\be
\!^4 R=e^{-\sigma}[R-\nabla^2\sigma-\frac{2}{{\cal D}}\nabla^2 {\cal D}
+ \frac{1}{2{\cal D}^2} (\nabla {\cal D})^2] + \frac{2}{{\cal D}},
\label{eq7}
\ee
where $\!^4 R$ is the 4D Ricci scalar, $R$ is the 2D Ricci scalar
associated with $g_{ab}$, and $\nabla$ is the
covariant derivative associated with $g_{ab}$. If we also note that
$\sqrt{-\!^4g}d^4x=\sqrt{-g}d^2x{\cal D}e^{\sigma} \sin \theta d\theta d\phi$,
we may rewrite the gravitational part of the action \rq{10} for $D=4$ as
\be
S = 4\pi \int d^2x \sqrt{-g} \left( {\cal D}(r)R
+ [\sigma'{\cal D}'+\frac{{\cal D}'^2}{2{\cal D}}] (\nabla r)^2
+2\Lambda e^{\sigma} \right),    \label{eq9}
\ee
where the $\theta$, $\phi$ integration has been carried out, the prime
denotes derivative with respect to $r$, and $\sigma \to \sigma +
\ln(\Lambda)$ for convenience.

If we take $V(r)= 2\Lambda e^{\sigma(r)}$ and
$H(r)=\sigma'{\cal D}'+{\cal D}'^2/2D$,
\rq9 is a special case of the general vacuum action for dilaton gravity
\cite{BanksL,Rtdilat}
\be
S=\int d^2x\sqrt{-g}(H(r) g^{ab}\nabla_{a}r\nabla_{b}r + V(r) + D(r)R).
\label{eq5}
\ee
In the absence of any matter action this model  actually only depends upon
the one function $V$,  since reparametrizations of the field $r$
accompanied by $r$-dependent Weyl rescalings of the metric allow one to
relate models with different $H$'s and $D$'s \cite{BanksL,Rtdilat}.
The action \rq5 reduces
to the gravitational part of (\ref{eq17}) for $H=\frac{1}{2}$, $D=r$ and
$\Lambda\to 0$, and to the effective target space action for
non-critical string theory \cite{MSW,EDW} for $D=e^{2r}=\frac{1}{4}H$.

We may see that this action contains the same information as \rq{10} by
noting that the field equations $\!^4 G_{\mu\nu} =0$ follow from it. $\!^4
G_{ab}=0$ follows from variation with respect to $g_{ab}$, and
\be
\!^4 G_{\phi\phi} = \sin^2(\theta) \!^4 G_{\theta\theta} =0      \label{neq1}
\ee
follows from variation with respect to $r$.

To summarize, we have shown that a rescaling of the gravitational coupling
constant by a factor of $(\frac{D}{2}-1)$ permits one to take a $D\to 2$
limit of general relativity. The resultant action, \rq{17}, yields a
2D theory in which stress-energy generates gravity and gravity acts in turn
upon stress-energy in a manner quite analogous to general relativity
\cite{Rsimp}. This theory is also found as a special case of 2D dilaton
gravity as derived by more tradition dimensional reduction methods. It
stands in contrast to the wide class of other dilaton gravity theories in
that the classical evolution of the gravity/matter system is unaffected by
the evolution of any Brans-Dicke scalar. In this sense the theory based on
\rq{17} may be said to be the closest thing there is to general relativity
in two dimensions.

\section*{Acknowledgments}
This work was supported by the Natural Sciences and Engineering Research
Council of Canada.

\end{document}